\documentclass[pra,twocolumn,showpacs,showkeys,preprintnumbers,amsmath,amssymb]{revtex4}

\usepackage{graphicx}
\usepackage{dcolumn}
\usepackage{bm,amsmath,amsfonts,amssymb,mathrsfs}
\newcommand{\ket}[1]{|#1\rangle}
\newcommand{\bra}[1]{\langle#1|}

\newcommand{\be}[0]{\begin{equation}}
\newcommand{\ee}[0]{\end{equation}}
\newcommand{\bea}[0]{\begin{eqnarray}}
\newcommand{\eea}[0]{\end{eqnarray}}

\newcommand{\Tr}[1]{\text{Tr}\left [ #1 \right]} 
\newcommand{\Tra}[2]{\text{Tr}_{#1}\left [ #2 \right]}

\usepackage[matrix,frame,arrow]{xy}
\usepackage{amsmath}
\newcommand{\qw}[1][-1]{\ar @{-} [0,#1]}
\newcommand{\qwx}[1][-1]{\ar @{-} [#1,0]}

\newcommand{\gate}[1]{*{\xy *+<.6em>{#1};p\save+LU;+RU **\dir{-}\restore\save+RU;+RD **\dir{-}\restore\save+RD;+LD **\dir{-}\restore\POS+LD;+LU **\dir{-}\endxy} \qw}

\newcommand{\control}{*!<0em,.025em>-=-{\bullet}}

\newcommand{\ctrl}[1]{\control \qwx[#1] \qw}

\newcommand{\qswap}{*=<0em>{\times} \qw}
\newcommand{\lstick}[1]{*!R!<.5em,0em>=<0em>{#1}}
\newcommand{\ustick}[1]{*!D!<0em,-.5em>=<0em>{#1}}

\newcommand{\Qcircuit}[1][0em]{\xymatrix @*[o] @*=<#1>}

\begin{document}
\title{Nonadditivity effects in classical capacities of 
quantum multiple-access channels}

\author{\L{}ukasz Czekaj}
\email{lczekaj@mif.pg.gda.pl}
\author{Pawe\l{} Horodecki}
\email{pawel@mif.pg.gda.pl}
 \affiliation{Faculty of Applied
Physics and Mathematics, Gda\'nsk University of Technology,
Gda\'nsk, Poland \\ 
National Quantum Information Centre of Gda\'nsk, 81-824 Sopot, Poland}

\date{\today}

\begin{abstract}
We study classical capacities of quantum multi-access channels in geometric terms 
revealing breaking of additivity of Holevo-like capacity. 
This effect is purely quantum since, as one points out, any classical multi-access 
channels have their regions additive. The observed non-additivity in 
quantum version presented here seems to be the first effect of this type with no 
additional resources like side classical or quantum information (or entanglement) involved.
The simplicity of quantum channels involved resembles butterfly effect in case of 
classical channel with two senders and two receivers.
\end{abstract}

\maketitle

\textbf{Introduction.} One of the central notions of quantum information theory is represented by 
quantum channels \cite{Science,BennettSeparation}. Many of them allow  to transmit quantum  information coherently 
\cite{Bennett,Barnum,BarnumForward} or transfer classical information, after suitable encoding 
into quantum states \cite{Holevo,SchumacherWestmoreland}. 
The corresponding notions of  quantum capacity $Q$ and classical capacity $C$ 
rise apparently hard open problems \cite{activation,Shor_additivity} of their additivity in biparty scenario
with especially problems related to $C$ attracting much attention recently (see \cite{Shor_additivity,Holevo1,Ruskai}).
On the other hand multiparty communication was analysed 
\cite{Winter,YardEtAlMultiple,HorodeckiEtAlNature,DCHnonadditivity,YardEtAlBroadcast,Demianowicz,QButterfly} 
with nonadditivity effects reported \cite{DCHnonadditivity,QButterfly} for analog of $Q$. 
However they require either supplementary resources like classical communication 
or have their classical analogs (see \cite{Butterfly}).  
Here we show that there are nonadditivity effects avoiding both the above features.
We provide specific examples of multiple-access channels and show how they exhibit nonadditivity 
of the classical capacity regions. 
This is purely quantum phenomenon since, as we point out, the corresponding 
regions for multiple-access classical channels \cite{CoverThomas} are always additive.
The revealed nonadditivity effects may shed new light on information transmission with help 
of quantum resources. They also constitute a natural arena for 
applications of all known techniques form bipartite channels.

\textbf{Capacity regions and geometric sum.}
Capacity region is a set of all rates achievable for channel.
For two channels $\Phi_1$, $\Phi_2$
and their the capacity regions ${\cal C}(\Phi_1)$ and ${\cal C}(\Phi_2)$ 
on defines a geometric (Minkowski) sum ${\cal C}(\Phi_1)+{\cal C}(\Phi_2)=\{\vec{u}_1 + \vec{u}_2:
 \ \vec{u}_1 \in  {\cal C}(\Phi_1) ,  \vec{u}_2 \in  {\cal C}(\Phi_2) \}$.
 The latter gives region of achievable rates in case when both channels are used
separately  ie. input states are not correlated across $\Phi_1,\Phi_2$ cut.
One  immediately  has ${\cal C}(\Phi_1)+{\cal C}(\Phi_2) \subseteq {\cal C}(\Phi_1 \otimes \Phi_2)$
since the inputs may be correlated. The converse inclusion defines additivity 
which - in case of one sender - one receiver scenario is a hard open issue \cite{Shor_additivity}.

Here we shall consider multi-access channel capacity region ${\cal C}$.
In quantum case of two senders and one receiver 
one defines the classical-quantum channel (cqc) state $\rho=\sum_{ij}p_i q_j e_i \otimes
 e_j \otimes \Phi( \varrho_i \otimes \varrho_j)$.  Here $e_i=|e_i\rangle \langle e_i|$ is a projector onto the standard basis element of classical part belongs to first (second) sender say Alice (Bob) while $\{ p_i, \varrho_i\}$, $\{ q_j, \varrho_j \}$ represent the ensambles of states send through the channel toward the receiver Charlie. Receiver is allowed to perform POVM measure to recover classical information encoded in quantum states. The capacity region ${\cal C}$ for given cqc state is described by \cite{Winter,Allachverdyan1}:
 \begin{eqnarray}
 &&R_A\leq I(A:C|B) \nonumber \\
 &&R_B\leq I(B:C|A) \nonumber \\
 &&R_A + R_B\leq I(AB:C)
\label{regions}
 \end{eqnarray}
$I(AB:C),I(A:C|B),I(B:C|A)$ can be viewed as {\it (conditional) mutual information} $I(AB:C)=S(\rho_A)+S(\rho_B)-S(\rho_{AB})$ ($I(A:C|B)=\sum_j p_j I(A:C|B=j)$)
of classical-quantum state shared between sender and receiver.

\begin{figure}[h]
	\centering
		\includegraphics[scale=0.6]{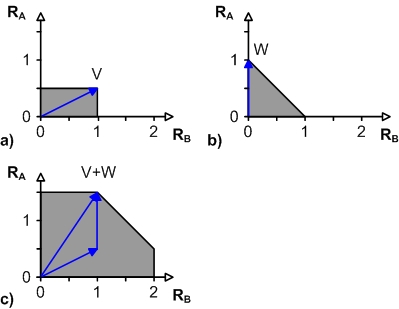}
	\caption{Here one has in turn the capacity regions of 
	a) classical single receiver messages from two independent binary symmetric channels with $H(p)=0.5$ 
	and $H(p)=0$ respectively b) classical XOR gate c) Minkowski sum of the two previous regions
	illustrating the additivity rule. 
	}
\label{fig:DiagramSumaObszarow}
\end{figure}

The formula naturally generalizes for more senders (see \cite{Winter}).
Quite remarkably it looks the same for classical channel 
\cite{CoverThomas} and then it can be shown to be additive (see Methods).
The Fig. \ref{fig:DiagramSumaObszarow} illustrates additivity of the regions for exemplary pair of classical channels.



\begin{figure}[htpb]
	\centering
		\includegraphics[scale=0.4]{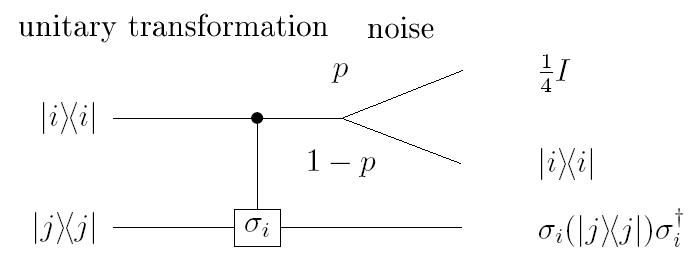}
	\caption{Circuit model of channel $\Phi^{p}$ with depolarizing noise. 
	The controlled Pauli matrices $\sigma_i\in\{I,\sigma_x,\sigma_y,\sigma_z\}$ are involved.}
	\label{fig:realnoisechannel}
\end{figure}

\textbf{Basic counterexample channel.}
Consider the case of two senders Alice and 
Bob and the following channel $\Phi^{p}$ that 
allows Alice to send a four level quantum system while 
Bob is supposed to send only one qubit system. Our model channel, 
is depicted schematically on Fig.  \ref{fig:realnoisechannel}. 

The capacity of the channel $\Phi^{p=1}$, can be easily found as follows.
Let us put partial trace instead of depolarization (both cases are 
completely equivalent ie. have no impact on the capacity regions).
Now if Alice sends fixed state, say $|0 \rangle$ then Bob message is not affected
which gives rise to the following rate vector $(R_A,R_B)=(0,1)$.
On the other hand if Bob sends fixed pure state, say $|0 \rangle$, then Alice may not affect it
sending $|0\rangle$ or may alter with Pauli matrix $\sigma_x$ by sending $|1\rangle$.
That case corresponds to the rate vector $(R_A,R_B)=(1,0)$.
Clearly sum of the rates cannot exceed one (since Charlie gets only one qubit). Thus, exploiting time sharing, 
we get the capacity region ${\cal C}(\Phi^{p=1})$:
\begin{equation}
R_{A} + R_{B} \leq 1
\end{equation}
We also introduce a trivial identity channel $\Psi^{id}$ that transmits ideally 
single qubits form Alice and Bob respectively \footnote{Here sending qubit from Alice 
is not needed but we introduce it for more natural geometrical visualization.}.
The capacity region of ${\cal C}(\Psi^{id})$ is:
\begin{eqnarray}
&&R_{A} \leq 1, \nonumber \\
&&R_{B} \leq 1
\end{eqnarray}

\begin{figure}[htbp]
	\centering
		\includegraphics[scale=0.6]{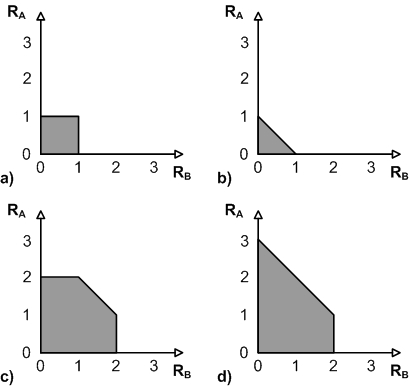}
	\caption{Illustration of a) identity channel capacity region $\Psi^{id}$  b) channel  $\Phi^{p=1}$ capacity region   c) Minkowski sum of the regions a) and b), d) capacity region of product of two channels	which is 
 greater than the sum c).}
	\label{fig:capregtoymodelall}
\end{figure}

Now we shall find the capacity region ${\cal C}(\Phi^{p=1} \otimes \Psi^{id})$.
The general idea is to explore the analog of dense coding \cite{DC}. Bob may send fixed maximally 
entangled state, say $|\Psi_{+}\rangle=\frac{1}{\sqrt{2}}
(|00\rangle +|11\rangle)$. Then Alice may alter it with her four states 
$|0\rangle,...,|3\rangle$ and send four independent messages to Charlie.
She may also send one additional bit by ideally transmitting part of $\Psi^{id}$.
This gives totally 3 bits of Alice rate ie. $(R_A,R_B)=(3,0)$.
Again, since Charlie gets three qubits, by  Holevo bound, 
sum of the  Alice and Bob rates can not exceed 3 bits. By the same argument 
Bob can not send more than two bits. Using the geometric sum of the previous 
regions gives finally the region ${\cal C}(\Phi^{p=1} \otimes \Psi^{id})$:
\begin{eqnarray}
&&R_{A}+ R_{B} \leq 3, \nonumber \\
&&R_{B} \leq 2
\end{eqnarray}
which is clearly grater than the geometric sum ${\cal C}(\Phi^{p=1}) + {\cal C}(\Psi^{id})$
as illustrated in Fig. \ref{fig:capregtoymodelall}.
This example which explores the kind of {\it remote dense coding} on Alice part
shows how easily that nonadditivity of capacity regions of different channels
may naturally occur in multiple-access channel.

For all $\Phi^p$ with $0<p<1$ we have $R_A<2$, hence one can observe that nonadditivity of capacity region occur also for $p<1$.

Note that in the case $\Phi^{p=1}$ we have single letter formulas for 
all the three capacities ie. entangled signals sent across 
inputs of {\it the same} channels will not help or 
- in other words - for all those channels we have 
${\cal C}^{(n)}(\Phi)\equiv \frac{1}{n}{\cal C}(\Phi^{\otimes n})$
is just equal to ${\cal C}(\Phi)$.
As we shall see subsequently this is not always true. 

\textbf{The presence of nontrivial noise: when single letter formula does not work.} 
Consider noisy version $\Phi^{p}$ with $p$ different from  zero or one.
We will show that in that case ${\cal C}^{(1)}(\Phi^{p})  \subsetneq {\cal C}^{(2)}(\Phi^{p})$.
Remarkably, the analysis will illustrate usefulness of tools from original bipartite additivity problem.
To estimate capacity region of the above channel even in single copy case seems to be 
not immediate. Thus we shall focus here on the maximal Alice transmission rate.

Following (\ref{regions}) the bound on Alice transmission rate in 
{\it single use} of the channel may be expressed by:
\begin{eqnarray}
&&R_A \leq \max \ \chi(\{p_i, \Phi(u_i \otimes v) \}= \nonumber \\ 
&&\max [S(\Phi(\sum_i p_i u_i \otimes v))
-\sum_ip_iS(\Phi(u_i \otimes v))]
\label{Alice-bound-1}
\end{eqnarray}
since the Holevo function can be always saturated on pure state ensambles. Maximum is taken over all Alice ensambles $\{p_i, u_i \}$ and all Bob pure states $\{ v \}$.
We shall prove below that this bound is tight and amounts to $\chi^{(1)}= H((2-p)/8,(2-p)/8,(2-p)/8,(2-p)/8,p/8,p/8,p/8,p/8)-
H(1-(3p/4),p/4,p/4,p/4)$. This can be seen from two facts: 
(i) the total entropy ie. the first term in the above bound 
is maximized by the Alice maximally mixed state (ii) all the terms 
in the second part $S(\Phi(u_i \otimes v))$ have the same minimum 
for $|u_i\rangle=|i\rangle$ (standard orthonormal basis). Hence follows that maximally mixed ensamble of Alice orthogonal states $\{ |i\rangle \}$ which at the same time maximizes the first and minimizes the second (averaged) term in (\ref{Alice-bound-1}) reaches the bound. 
The fact (i) and (ii) are proved in Methods. 

Consider now the case when we have {\it two uses} of the channel
$\Phi^{p}\otimes \Phi^{p}$ and Bob sends just maximally entangled state $|\psi_{+}\rangle$ while Alice sends just products of two 
maximally mixed ensambles like the one used before. The achieved Alice rate $\chi'^{(2)}$ \footnote{Prime stands here to stress that it is not optimized Holevo function ie. achieved for specific protocol.}
can be easily computed as a Holevo function of the ensamble of sixteen states 
$\Phi(e_i \otimes e_j \otimes \psi_+)$ with equal probabilities
and it amounts to $\chi'^{(2)}=- \big( \frac{3}{8} (2-p)p \log_2\frac{1}{64} (2-p)p 
+ \frac{1}{8} (4-6p+3p^2)\log_2\frac{1}{64} (4-6p+3p^2)
\big) - H(1-(3p/4),p/4,p/4,p/4)$ where the first term contributes to superadditivity.

\begin{figure}[htbp]
	\centering
		\includegraphics[scale=0.45]{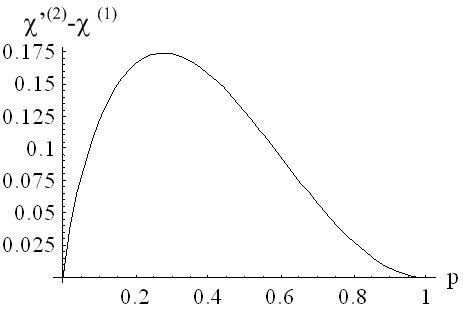}
	\caption{Difference between Alice's Holevo-like capacity for entangled and product coding.}
	\label{fig:realcapcomp}
\end{figure}

On Fig. \ref{fig:realcapcomp} the difference $\chi'^{(2)}-\chi^{(1)}$ 
is depicted showing that maximal possible rate of sending information by Alice
which implies nonadditivity ${\cal C}^{(1)}(\Phi^{p})\subsetneq {\cal C}^{(2)}(\Phi^{p})$
for any nontrivial $p$.
In particular the application of time sharing strategy implies that 
the triangle bounded by the three lines $y=-\chi'^{(2)}x+1$, $y=\chi^{(1)}$ and 
$x=0$ belongs to ${\cal C}^{(2)}(\Phi^{p})$ and not to ${\cal C}^{(1)}(\Phi^{p})$.

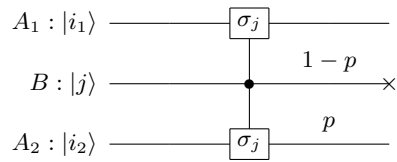
\begin{figure}[h]
	\centering
	\[\Qcircuit @C=0.8cm @R=.6cm {
\lstick{A_1 : \ket{i_1}}&\qw & \gate{\sigma_j} & \qw & \qw \\
\lstick{B : \ket{j}} & \qw & \ctrl{-1}\qwx[1] & \ustick{1-p} \qw & \qswap \\
\lstick{A_2 : \ket{i_2}} & \qw & \gate{\sigma_j} & \ustick{p} \qw & \qw
}\]			
	\caption{Circuit model of channel $\Gamma^p$. Up unitarity (between $A_1$ and $B$) occur with probability $1-p$ and down with $p$. The cross sign stands for partial trace.}
	\label{pic:three-input-toymodel}
\end{figure}

\textbf{Three sender channel with broken additivity.}
Here we shall consider another type of multiaccess channel 
with three senders site $A_1$, $A_2$ and $B$ and one receiver $C$.
The senders $A_1$ and $A_2$ send qubits while $B$ sends four-level system.
The channel is depicted on Fig. \ref{pic:three-input-toymodel}.

We shall consider configuration similar to one presented for case $\Phi^{p=1}$. We introduce trivial identity channel $\Psi_{id}$ that transmits ideal single qubits from $A_1$ and $A_2$ to receiver. For the case (i) when 
$A_1$ and $A_2$ sends single selected product state we immediately get: $R^{(1)}_B\leq 1$. When we allowed (ii) entanglement between many uses of $\Gamma^p$, the (regularized) rate $R_B$ will be bounded by $1.81$ (for details see Methods) and in the case when (iii) senders $A_1$, $A_2$ sends Bell state ($\ket{\Psi_{+}}$) where the first qubit is sent through channel $\Gamma^p$ and second through channel $\Psi^{id}$ transmition rate $R^{ent}_B$ becomes $2$. We again found situation when nonadditivity occur. (i) follows from Holevo bound for lines $A_1$ and $A_2$ and fact that channel where we know which unitarity (up or down) was performed has bigger capacity than $\Gamma^p$. (iii) we get immediately by superdense coding.
Note that by numerical analysis we also found for channel $\Gamma^{p=0.5}$ for (i), (iii) that even if $B$ sends with maximal rate, 
$A_1$ and $A_2$ can also achieve non zero rate.

\textbf{Conclusions} 
We have provided constructions of multiple-access channels that exhibit nonadditivities 
of classical capacity regions. 
First they are nonadditive in the sense 
that ${\cal C}^{(1)} \subsetneq {\cal C}^{(2)}$ ie., entanglement 
across two inputs of the same channel helps. This effect known in case 
of bipartite quantum capacities (due to nonadditivity of coherent information) 
is conjectured not to hold in classical biparty capacity.
Even more striking, unlike bipartite channel capacities \cite{Additivity-with-identity}, the presented capacity regions 
break additivity rule if supplied with identity channel. As one points out 
both types of nonadditivity have no classical analog.
%
The results seem to be the first examples of nonadditivity of capacities 
where (i) no additional resources are involved (ii) classical analogs are additive. 
It is also worth to note that minimal output entropy for presented channel is achieved for product states. We owe nonadditivity to growth of output variety due to effects associated with quantum dense 
coding \cite{DC}.
Cumulative rate ($R_A+R_B$) is still additive. On the other hand the results show that multiparty 
channel scenarios may be arena for efficient exploiting some of tools known from bipartite case. 
The simplicity of our initial channel 
that breaks additivity with identity channel resembles to some extend the classical 
butterfly-effect with two senders and two receivers \cite{Butterfly}.
Note that one can ask the same question for other type of multiuser scenarios. 
It is also interesting how the entanglement assisted classical 
capacity will behave with respect to additivity since naive 
extension of our approach to that case does not work. Also the analysis of presented 
effects for more complicated noise models and in continuous variables domain is an interesting 
problem but it will be considered elsewhere. Finally one may hope that the present 
work will stimulate general research on the role of dense coding in quantum networks.

%

\section{Methods}
\textbf{Additivity of multiple-access classical capacity regions.}
For any quantum multiple-access channel $p(y|x_1,\ldots,x_n)$
the capacity region is determined by 
the following set of inequalities \cite{CoverThomas}:
\begin{equation}
R(S)\leq I(X(S):Y|X(S^C))
\label{classical-capacity}
\end{equation}
parametrized by $S$ representing all possible subsets of senders $S \subseteq \{ X_1,...,X_n \}$.
$S^{C}$ stand for complements of $S$ and $R(S)$ are sums of transmission rates 
$R(S)=\sum_{X_i\in S}R(X_i)$ of senders $X_i \in S$ 
to the single receiver $Y$.
Consider now the classical channel being the product of two other channels:
$p(y|x_1,\ldots,x_n)=p(y_1|x_{1,1},\ldots,x_{1,n})p(y_2|x_{2,1},\ldots,x_{2,n})$.
Again one considers the bound on the sum of rates where $S=S_1 \cup S_2$ 
represents now the subset of $\{ X_{1,1},...,X_{n,1},X_{1,2},...,X_{n,2} \}$
with $S_1$, $S_2$ being the subsets of the first (second) group of receivers.
The following inequality
(which we leave as an exercise for the reader) 
\begin{eqnarray}
&&I(X(S):Y|X(S^C))
\leq I(X(S_1):Y_1|X(S_1^C)) +\nonumber \\
&&+I(X(S_2):Y_2|X(S_2^C))
\end{eqnarray}
clearly proves the geometric additivity of the capacity regions 
since capacity regions of the channels treated separately are just determined 
by the inequalities $R(S_1)\leq I(X(S_1):Y_1|X(S_1^C))$, 
$R(S_2)\leq I(X(S_2):Y_1|X(S_2^C))$. Note that this ensures in particular 
that multiple-access capacity region (\ref{classical-capacity})
is of single letter form.

\textbf{Finding maximum of total entropy in case of channel $\Phi^{p}$.}
Here we shall show that maximum of $S(\Phi^{p}(\rho \otimes v))$ 
is reached by $\rho=I/4$. First observe that the considered entropy 
can be seen as a concave function of  state $\rho$
so it is enough to prove that the entropy has a critical point ie. its derivative along any 
(traceless) direction $\Delta$ vanishes at $\rho=I/4$.  
We use the following formula \cite{Shorcalc}:
\begin{equation}
\left.\phantom{|}
\frac{\partial S(\varrho + \alpha \delta)}{\partial \alpha} \right|_{\alpha=0} =  - \Tr{\delta\log\varrho}
\label{pochodna}
\end{equation}
for any traceless $\delta$ and quantum state $\varrho$. Now we put into that formula 
$\varrho=\Phi^{p}( \frac{1}{4}\text{I} \otimes v)=(1-p)U(\frac{1}{4}\text{I}\otimes\ket{v}\!\bra{v})U^\dagger+
  p \frac{1}{8} \text{I}\otimes\text{I} $
and 
$\delta=\Phi^{p}(\Delta \otimes v)=(1-p)U(\Delta \otimes v)U^\dagger+
  p \frac{1}{4}\text{I}\otimes\Tra{A}{ U (\Delta \otimes v )U^\dagger }$  
with $U=\sum_i e_i \otimes \sigma_i$.
Defining the vectors four-dimensional projector $P= U (I \otimes  v) U^{\dagger} $ one finds the operator $log \varrho = log(p/8) (I \otimes I - P) + log((2+p)/4)P$. Then one proves vanishing of (\ref{pochodna}) 
via sequence of not difficult, though tedious calculations which will 
be presented elsewhere.

\textbf{States with minimal output entropy in case of channel $\Phi^{p}$.}
There is a theorem \cite{KingDep} 
saying that minimal output entropy of tensor product of depolarizing channel with identity channel 
is saturated by product pure states. In case of four dimensional depolarizing channel 
product with ideal qubit channel this is equal to $H(1-(3p/4),p/4,p/4,p/4)$
and can be, in particular, achieved by any arbitrary pure state of the form 
$|u\rangle|v'\rangle$. Since our channel $\Phi^{p}$ is a
product channel (composed of depolarizing channel and identity) follows the entangling unitary  operation we shall achieve the average minimum output entropy only if 
we put an input state that unitary operation will transform into
a product state of the form $|u\rangle|v'\rangle$ (since any entangled state 
is not better by the theorem mentioned above). 
The latter is produced in particular by any of the four inputs $|i\rangle|v\rangle$ with $v$ again arbitrary. Taking 
all the four inputs of the latter form we minimize the average entropy (second term in (\ref{Alice-bound-1})).

\textbf{Regularized maximal rate $R_B$ for channel $\Gamma^p$.}

We shall estimate $R_B^{(n)}$. Message $B$ may be always chosen to be in the standard basis since the receiver output is invariant under  the von Neumann measurement in standard basis on system $B$ that proceeds action of $\Gamma^p$.
Therefore we may simulate channel $\Gamma^p$ as a classical channel $\Lambda^p:B\mapsto B_1\otimes B_2$ followed by unitary operation $U=U_1\otimes U_2$ where $U_i$ depends on (classical) value of 
$B_i$ and acts further on subsystem $A_i B_i$ ($i$ specifies sender $A_1$ or $A_2$). After action of the unitary operation we trace out subsystem $B_1B_2$. Channel $\Lambda$ maps $i\in B$ to $(0,i)$ with probability $p$ and with probability $1-p$ we get $(i,0)$.
Suppose now we have $n$ copies of $\Gamma^p$ at our disposal.
Sender $A_{i}$ is allowed to prepare any state 
$\ket{\Psi_{A_i}}\in A^{\otimes n}_{i}$ on his subsystem. At the same time sender $B$ sends random vector variable $(b^1,\ldots,b^n)$ which is mapped to $(b^1_1,\ldots,b^n_1)(b^1_2,\ldots,b^n_2)$.
Assume now that by appropriate choose of $\ket{\Psi_{A_i}}$ we can perform any coding:
\begin{equation}
E_i:B_i^n\ni (b^1_i,\ldots,b^n_i) \mapsto \rho^{(b^1_i,\ldots,b^n_i)}_i\in A_i^{\otimes n}
\end{equation}
Receivers gets state $\rho_1\otimes\rho_2$ and performs on it POVM to get maximum information about $B^n$. Result of POVM is recorded in $\hat{B}^n$.
We shall denote: $H(B^n|\hat{B}^n) = n\epsilon_n$ (we do not assume that transition is perfect). $R_B^{(n)}$ can be expressed as:
\begin{equation}
n R_B^{(n)}\leq \max_{p(B^n)}I(B^n:\hat{B}^n) = \max_{p(B^n)}H(B^n)-n\epsilon_n
\label{eq:R_B_Def}
\end{equation}
Measurement correspond to quantum operation. That allows us to write following inequalities:
\begin{equation}
H(B^n|\rho_1\rho_2) \leq H(B^n|\hat{B}^n) = n\epsilon_n \label{eq:epsconst}
\end{equation}
Assume receiver obtains complete state $\rho_2$. It leads us to following estimation of $n\epsilon_n$ depending on $p(B)$:
\begin{eqnarray}
n &\geq& S(\rho_1)\\
&\geq&S(\rho_1|\rho_2) \\
&=&S(\rho_1|\rho_2) + S(B^n|\hat{B}^n) - S(B^n|\hat{B}^n) \\
&\geq&S(\rho_1|\rho_2) + S(B^n|\rho_1,\rho_2) - n \epsilon_n \label{eq:fano}\\
&=&S(B^n \rho_1|\rho_2) - n \epsilon_n \label{eq:chainrule}\\
&\geq&S(B^n \rho_1|B^n_2) - n \epsilon_n \label{eq:quantop}\\ 
&\geq&H(B^n|B^n_2) - n \epsilon_n \label{eq:classcorr}
\end{eqnarray}
where we use facts: (\ref{eq:fano}) follows from (\ref{eq:epsconst}), (\ref{eq:chainrule}) follows from chain rule, encoding is quantum operation (\ref{eq:quantop}), $\rho_1$ and $B^n$ are correlated classically (\ref{eq:classcorr}). Because $B^n_2$ depends only on $B^n$ we can express $H(B^n|B^n_2)$ in terms of $p(B^n)$ and parameter $p$.
Combining (\ref{eq:classcorr}) and (\ref{eq:R_B_Def}) we get:
\begin{eqnarray}
R_B^{(n)}&\leq&\max_{p(B^n)}\frac{1}{n}\left( H(B^n)-H(B^n|B_2^n)+n\right)\\
&=&\max_{p(B^n)}\frac{1}{n}\left(I(B^n:B^n_2)+n\right)\\
&=&\max_{p(B)}I(B:B_2)+1
\end{eqnarray}
Without loss of generality we can assume that $p\leq 0.5$.
Therefore by numerical calculation we get $R_B^{(n)} < 1.81$ for all $p\leq 0.5$.
Result is independent on $n$ hence we get finally that maximal regularized 
rate of Bob transmission  is bounded by $1.81$. 

\acknowledgments This work was prepared under the the EU integrated project SCALA.
We thank \L{}ukasz Pankowski for drawing our attention to classical butterfly effect.
For numerical calculations we used the following packages:
	GSL - GNU Scientific Library v1.9
	 and CGAL - Computational Geometry Algorithms Library v3.3.1 .

\end{document}